\newcommand{\be}{\begin{equation}}\newcommand{\ee}{\end{equation}}
\newcommand{\bea}{\begin{eqnarray}}\newcommand{\eea}{\end{eqnarray}}
\newcommand{\nn}{\nonumber}\newcommand{\p}[1]{(\ref{#1})}
 \newcommand{\lb}[1]{\label{#1}}
\newcommand\bt{\bar{\theta}}
\newcommand\th{\theta}
\newcommand\tp{\theta^+}
\newcommand\btp{\bar{\theta}^+}
\newcommand\tpm{\theta^{\mu+}}
\newcommand\tpa{\theta^{\alpha+}}
\newcommand\btpm{\bar{\theta}^{\dot{\mu}+}}
\newcommand\btpa{\bar{\theta}^{\dot{\alpha}+}}
\newcommand\btpn{\bar{\theta}^{\dot{\nu}+}}
\newcommand\tm{\theta^-}
\newcommand\btm{\bar{\theta}^-}
\newcommand\tma{\theta^{\alpha-}}
\newcommand\tmm{\theta^{\mu-}}
\newcommand\btma{\bar{\theta}^{\dot{\alpha}-}}
\newcommand\btmm{\bar{\theta}^{\dot{\mu}-}}
\newcommand\btmn{\bar{\theta}^{\dot{\nu}-}}

\newcommand\pmn{\partial_{\mu\dot{\nu}}}
\newcommand\ada{{\alpha\dot{\alpha}}}
\newcommand\adb{{\alpha\dot{\beta}}}
\newcommand\mdn{{\mu\dot{\nu}}}
\newcommand\ap{{\alpha +}}
\newcommand\dap{{\dot{\alpha}+}}

\newcommand\ha{\widehat{\alpha}}
\newcommand\hb{\widehat{\beta}}
\newcommand\hm{\widehat{\mu}}
\newcommand\hn{\widehat{\nu}}
\newcommand\Hm{\widehat{m}}
\newcommand\Hn{\widehat{n}}

\newcommand\s{\scriptscriptstyle}
\newcommand\A{\s A}
\newcommand\B{\s B}
\newcommand\M{\s M}
\newcommand\N{\s N}
\newcommand{\pp}{\s ++}
\newcommand{\m}{\s --}
\newcommand{\0}{\s 0}

\newcommand{\5}{\s 5}
\newcommand{\pT}{\s(+3)}
\newcommand{\Dp}{D^{\pp}}
\newcommand{\Dm}{D^{\m}}
\newcommand{\Dep}{\Delta^{\pp}}
\newcommand{\Dem}{\Delta^{\m}}
\newcommand{\dpp}{\partial^{\pp}}
\newcommand{\dm}{\partial^{\m}}
\newcommand{\Dap}{D^+_\alpha}

\newcommand{\bDap}{\bar{D}^+_{\dot{\alpha}}}

\newcommand\dza{d\zeta^{\s(-4)}}

\documentstyle[12pt]{article}

\begin{document}
\renewcommand{\thefootnote}{\fnsymbol{footnote}}
\begin{titlepage}
\hfill hep-th/9803202\\

\hfill JINR E2-98-67\\

\hfill March 1998 \vspace{1cm} \\

\begin{center}
{\bf B.M.Zupnik}\footnote{On leave of absence from:
Institute of Applied Physics, Tashkent State University}
\vspace{0.5cm} \\
{\it Bogoliubov Laboratory of Theoretical Physics, Joint Institute for
Nuclear Research, 141980, Dubna, Moscow Region, Russia}\\
E-mail: zupnik@thsun1.jinr.dubna.su
\vspace{0.5cm}

{\large\bf BACKGROUND HARMONIC SUPERFIELDS IN N=2 SUPERGRAVITY}
\vspace{1cm} \\
\end{center}

 A modification of the harmonic superfield formalism in  $D=4, N=2$
supergravity using a subsidiary condition of covariance under the
background supersymmetry with a central charge ($B$-covariance) is
considered. Conservation of analyticity together with the $B$-covariance
leads to the appearance of linear gravitational superfields. Analytic
prepotentials arise in a decomposition of the background linear
superfields in terms of spinor coordinates and transform in a nonstandard
way under the background supersymmetry. The linear gravitational
superfields can be written via spinor derivatives of nonanalytic spinor
prepotentials. The perturbative expansion of the extended supergravity
action in terms of the $B$-covariant superfields and the corresponding
version of the differential-geometric formalism are considered. We discuss
the dual harmonic representation of the linearized extended supergravity,
which corresponds to the dynamical condition of Grassmann analyticity. \\
\vspace{0.5cm}

\end{titlepage}

\renewcommand{\thefootnote}{\arabic{footnote}}
\setcounter{footnote}0
\setcounter{equation}0
\section{Introduction}
\hspace{0.5cm}
Interactions of  physical fields of the extended supergravity ($SG^2_4$):
graviton $g_{mn}$, gravitino $\psi^{\alpha i}_m$ and the abelian gauge
field $A_m$ have firstly been investigated in the framework of the
formalism that takes into account equations of motion for the proof of
local supersymmetry \cite{FSN,FSZ}. The problem of auxiliary fields of
$SG^2_4$ has been solved in refs. \cite{FV,WHP}. We shall consider a
 version of the Einstein $N=2$ supergravity with 40 boson and 40 fermion
field components. It should be noted that the component tensor calculus of
 $SG^2_4$ leads to very tedious calculations in constructing  interactions
with matter supermultiplets and analyzing  quantum properties of the
theory. So  various superfield approaches to the study of the extended
supergravity have been developed intensively in parallel with the
component analysis \cite{BS,SW,CNG,So,GS}.

A geometric approach to the superfield description of $SG^2_4$ has been
proposed in the method of harmonic superspace ($HSS$) \cite{GIK1}.
Analytic prepotentials of $SG^2_4$ appear in a decomposition of the
invariant harmonic derivative $\Dep$ in terms of the operators of partial
derivatives in $HSS$. We have shown in \cite{Z1} that harmonic superfield
constraints of zero dimension in $SG^2_4$ can be solved in the special
{\it flat} coordinates by analogy with the solution of the corresponding
harmonic constraints of the nonabelian supergauge theory \cite{GI2,Z2,Z4}.
Using this solution the authors of ref. \cite{GNS} have solved all
superfield constraints of $SG^2_4$ and  constructed the superfield action
nonlinear in prepotentials. The conformal $N=2$ supergravity and
alternative versions of the extended Einstein $SG$ have also been studied
in $HSS$ \cite{GIOS}.

Effectiveness of superfield formulations in supersymmetric theories is
mainly connected  with  simplification of quantum calculations. It should
be underlined that a procedure of $HSS$-quantization of the extended
supergravity seems us technically more complicated than the quantization
of Yang-Mills theory in $HSS$ \cite{GI2}. First of all, $SG^2_4$ has
several analytic prepotentials analogous to one matrix prepotential of the
gauge theory. Besides, in refs. \cite{GIK1,GNS}, the gauge symmetry of the
classical formalism of the extended supergravity has been considered, and
the additional background supersymmetry of perturbative expansion of the
$SG^2_4$ action in degrees of the gravitational constant $\kappa$ has not
been discussed there.  We know from the formalism of quantization of the
superfield $N=1$  supergravity \cite{GGRS,ZP} that the background
supersymmetry is very important in  constructing  superfield Feynman
rules. This work develops the investigation of the
background-supersymmetry problem in the harmonic formalism of $SG^2_4$
started in \cite{ZT}.

In sect. \ref{B}, we discuss a connection between different bases of
differential operators in the background $HSS$ and show that gravitational
superfields depending linearly on the spinor nonanalytic coordinates $\tm$
(linear harmonic superfields) naturally arise  in the covariant basis.
Analytic prepotentials of a holonomic basis \cite{GIK1} can be treated in
this approach as coefficients of the $\tm$-decomposition of background
linear superfields; so they have nonstandard transformations  with respect
to the background supersymmetry.

Note that the use of an analytic compensator \cite{GNS} guarantees only
the background supersymmetry without the central charge in the $SG^2_4$
action, but the formal covariance is hidden after the shift of this
compensator superfield on the flat part manifestly depending on spinor
coordinates.

In sect. \ref{H}, we consider a solution of the analyticity condition in
$SG^2_4$  which guarantees the covariance with respect to the background
supersymmetry with the central charge ($B$-co\-va\-ri\-ance).  Alternative
possibilities of  choosing unconstrained superfield variables in the
harmonic superspace are studied. In particular, we discuss the solution of
the $SG^2_4$ constraints in $HSS$ corresponding to the representation of
linear gravitational superfields through the unconstrained harmonic spinor
superfield of dimension $d=5/2$. In a special gauge this solution can be
expressed in terms of the harmonic-in\-dependent spinor prepotential,
which has been considered earlier while describing the linearized
supergravity in the ordinary superspace \cite{GS}. $B$-\-co\-vari\-ant
solutions of the $SG^2_4$ constraints are discussed in sects. 4 and 5.

Sect.\ref{D} is devoted to a discussion of terms  in the harmonic $SG^2_4$
action quadratic in the vector and scalar superfields. The structure of
these terms is similar to the structure of the action in  $N=2$ gauge
theory \cite{ZT}. Note that the quadratic action has an additional global
symmetry. $B$-co\-va\-riant decomposition of the $SG^2_4$ action in
$\kappa$ can be constructed by an iteration method taking into account the
gauge invariance.

In ref.\cite{GNS}, the nonlinear action of $SG^2_4$ has a form of the
 action for an analytic compensator in which the flat part manifestly
depending  on spinor coordinates is separated. This representation allows
one to prove the gauge invariance; however, it does not possess the
manifest $B$-co\-va\-riance.

In sect.\ref{F}, we discuss  the alternative harmonic formalism of the
linearized $SG^2_4$, in which spinor prepotentials of dimension $d=3/2$
determine the dual invariant harmonic derivative $\Dem$, a spinor
component of the torsion with $d=-1/2$ vanishes identically, and
equations of motion are equivalent to the dynamical analyticity condition
(zero-curvature representation ). This formalism is constructed by analogy
with a dual formulation of the $N=2$ Yang-Mills theory \cite{Z3}. We hope
that superfield methods will help us to study the quantum structure of
$N=2$ supergravity.

In Appendix, the definitions and notation of the basic derivatives in the
flat $HSS$ and some other useful formulae from refs. \cite{GIK1,GI2} are
written down.

\setcounter{equation}{0}
\section{\label{B} FLAT BACKGROUND HARMONIC \newline SUPERSPACE }
\hspace{0.5cm}
It is convenient to consider superfields with a real central charge in the
harmonic superspace $HSS(Z)$ \cite{GIK1} with the coordinates $u^\pm_i$
and
\be
z^{\M}_{\A}=(x^{\5}_{\A},~x^m_{\A},~\tpm,~\btpm,~\tmm,~\btmm)~,\label{B1}
\ee
 where $u^\pm_i$ are the $SU(2)/U(1)$ harmonics ,  $x^{\5}_{\A}$ is
a special coordinate associated with the central charge  $Z$, and $m, \mu,
\dot{\mu}$ are vector and spinor indices of the Lorentz group $SL(2,C)$.
 The analytic subspace  $ASS(Z)$ is defined by the coordinates
\be
\zeta(Z)=(x^{\5}_{\A},~\zeta)=( x^{\Hm}_{\A},~\theta^{\hm +})~,\label{B2}
\ee
where $\zeta$ describes 4-dimensional analytic coordinates, and the
notation $\Hm=(5,~m)$ and $\hm=(\mu,~\dot{\mu})$ is introduced.

An introduction of the fifth coordinate is connected with a geometric
interpretation of the central charge, and superfields can only have the
cyclic dependence on $x^{\5}_{\A} $
\be
\frac{\partial}{\partial x^{\5}_{\A}}\Phi =iZ \Phi~. \lb{B2b}
\ee
The gauge supergroup of the extended supergravity   $\Lambda SG^2_4 $
is defined quite naturally using the transformations of the
analytic coordinates $ z^{\s M}_{\A} $ \cite{GIK1}.

The basic infinitesimal parameters of the gauge transformations
do not depend on $x^{\5}_{\A},~\tm,~\btm$
\bea
&&\delta x^{\Hm}_{\A}=\lambda^{\Hm}(\zeta,u)\lb{B3}~,\\
&&\delta \theta^{\hm+}=\lambda^{\hm+}(\zeta,u)~.
\lb{B4}
\eea
Harmonics do not transform in $\Lambda SG^2_4$ , and transformations of
 spinor coordinates with the charge $-1$ depend on all $HSS(Z)$
coordinates besides $x^{\s5}_{\A}$
\be
\delta \theta^{\hm-}=\lambda^{\hm-}(x^m_{\A},\th^\pm,\bt^\pm,u)~.
 \label{B5}
\ee

A local gauge transformation of the general scalar superfield
$\Phi(z_{\A},u)$ in $\Lambda SG^2_4$
\be
\hat{\delta}\Phi=-\Lambda \Phi
 \label{B6}
\ee
is defined by the transformation operator  $\Lambda$ that includes the
 analytic operator $\lambda$
\be
\Lambda=\lambda +\lambda^{\hm-}\partial^+_{\hm}~,\quad \lambda=
\lambda^{\Hm} \partial_{\Hm}^{\A}+
\lambda^{\hm+}\partial^-_{\hm}~. \label{B7}
\ee

Gauge transformations of superfields preserve the Grassmann analyticity
and the cyclicity condition \p{B2b}
\be
[ \partial_{\hm}^+ , \lambda ]=0~,\quad[ \partial_{\hm}^+ , \Lambda ]=
(\partial_{\hm}^+\lambda^{\hn-})\partial_{\hn}^+ ,\quad
[ \partial_{\5},\Lambda ]=0~. \label{B8}
\ee
The local gauge transformations of analytic superfields
have the form $\hat{\delta}\omega=-\lambda \omega$.

 The differential operator  $\Lambda$ in eq.\p{B7} is given in
the holonomic basis, i.e. in a form of decomposition in terms of partial
derivatives $\partial^{\A}_{\M}$. Gravitational prepotentials of
 $SG^2_4$ have also been defined  in this holonomic basis as coefficients
 of the $\Lambda$-in\-va\-riant harmonic differential operator
\be
\Dep=\dpp + H^{\Hm\pp}\partial^{\A}_{\Hm}+H^{\hm\pT}\partial^-_{\hm}
+H^{\hm+}\partial^+_{\hm} ~. \label{B9}
\ee

Full gauge variations of these prepotentials have a very simple form
\be
\delta_\lambda H^{\Hm\pp}=\Dep \lambda^{\Hm}~,\quad \delta_\lambda
H^{\hm\s(\pm1+2)}=\Dep \lambda^{\hm\s\pm}~.\lb{Htr}
\ee
Commutator of two full variations can be calculated very simply in virtue
of the invariance of the operator
$\Dep$
\bea
&&[\delta_{\s2},\delta_{\s1}]H^{\M\pp}=\Dep\lambda^{\M}(1,2)~,
\lb{commu1}\\
&&\lambda^{\widehat{\M}}(1,2)=
\lambda_{\s2}\lambda^{\widehat{\M}}_{\s1}-
\lambda_{\s1}\lambda^{\widehat{\M}}_{\s2}~, \lb{commu}\\
&& \lambda^{\hm-}(1,2)=
\Lambda_{\s2}\lambda^{\hm-}_{\s1}-
\Lambda_{\s1}\lambda^{\hm-}_{\s2}\lb{commu2}~,
\eea
where $\widehat{\scriptstyle M}=(\Hm,\hm+)$ and the differential operators
 $\Lambda$ and $\lambda$ \p{B7} are used.

We shall use the following gauge for the nonanalytic gauge transformations
 \cite{GIK1}:
\be
\Dep\lambda^{\hm-}=\lambda^{\hm+}~. \label{B10}
\ee
The matrix of induced tangent transformations is covariantly independent
of harmonic variables in this gauge  $\Dep\partial_{\hm}^+
\lambda^{\hn-}=0$.

 The condition \p{B10} corresponds to the following gauge of the
nonanalytic prepotential:
\be
H^{\hm+}=\th^{\hm+}~. \lb{B12}
\ee

 In ref. \cite{GIK1}, a flat limit of gravitational superfields is
defined and a possibility of expansion in terms of the gravitational
constant $\kappa$ with respect to this limit is discussed. We shall
consider the flat superspace  $HSS(Z)$ as the background classical
superspace of $N=2$ supergravity. The background supersymmetry with the
real central charge will be denoted by a symbol $B_4^2(Z)$, and the
background superfields will be called $B$-su\-per\-fields. Gravitational
and matter superfields and the interaction in any degree in $\kappa$
should be covariant under $ B_4^2(Z)$ in this approach.

It should be underlined that the holonomic basis of  $HSS(Z)$ is
noncovariant with respect to  $B_4^2(Z)$, and the corresponding
prepotentials $H^{\Hm\pp}$ \p{B9} cannot be treated  as
$B$-su\-per\-fields.

 In ref. \cite{Z1}, the real coordinates have been used in $HSS(Z)$
\be
z^{\M}=(x^{\5},~x^m,~\th^\mu_i,~\bt^{i\dot{\mu}}) \lb{B13}
\ee
and the $B$-co\-va\-riant decomposition of the invariant analytic
operator has been considered
\be
\Dep=\dpp + G^{\pp}~,\quad [D^+_{\ha},G^{\pp}]=0~,\quad
G^{\pp}=\hat{h}^{\M\pp}\partial_{\M}\equiv h^{\M\pp}\partial_{\M}^{\A}~.
 \lb{B14}
\ee

This representation is useful for studying an iterative solution
 to the harmonic equations of $SG^2_4$; however, it also uses the
holonomic bases  $\partial_{\M}$ and $\partial_{\M}^{\A}$ noncovariant
with respect to $B^2_4(Z)$. Gauge transformations of the prepotentials
$h^{\M\pp}$ and $\hat{h}^{\M\pp}$ can readily be obtained from the
transformations \p{Htr} by using the relations
\bea
&\hat{h}^{m\pp}=h^{m\pp}=H^{m\pp}+2i\tp\sigma^m\btp~,\quad
 h^{\hm\pT}=H^{\hm\pT}~,\quad h^{\hm+}=0~,&\lb{apr1}\\
&\hat{h}^{\5\pp}=h^{\5\pp}=H^{\5\pp}+i(\btp)^2-i(\tp)^2
~,\quad \hat{h}^{\hm k\pp}=
u^{k-}h^{\hm\pT}~.&\lb{apr2}
\eea
Note that these prepotentials have a dimension $d=1$ and $1/2$, they do
not contain flat parts and are proportional to the constant $\kappa$.

It is convenient to study component stuff of the analytic prepotentials
in the physical  $WZ$-gauge \cite{GIK1}
\bea
&&h^{m\pp}_{\s WZ}=\kappa[-2i\tp\sigma^a\btp h_a^m(x_{\A}) +
(\btp)^2\tpa u^-_k\psi^{mk}_\alpha(x_{\A})  +\nn\\
&&+(\tp)^2\btpa u^-_k\bar{\psi}^{mk}_{\dot{\alpha}}(x_{\A})
+(\tp)^2(\btp)^2 u^{k-} u^{l-} V_{(kl)}^m(x_{\A})]~,\lb{comp1}\\
&&h^{\5\pp}_{\s WZ}=\kappa[i\tp\sigma^a \btp A_a(x_{\A}) +(\btp)^2\tpa
 u^-_k\rho^k_\alpha(x_{\A})+\nn\\
&&+(\tp)^2\btpa u^-_k \bar{\rho}^k_{\dot{\alpha}}(x_{\A})+
(\tp)^2(\btp)^2 u^-_k u^-_l S^{(kl)}(x_{\A})]~,\lb{comp2}\\
&&h^{\mu\pT}_{\s WZ}=\kappa\{ (\tp)^2\btp_{\dot{\alpha}}
[B^{\mu\dot{\alpha}}(x_{\A})+iC^{\mu\dot{\alpha}}(x_{\A})]+
(\btp)^2\tp_\alpha [\varepsilon^{\mu\alpha}(M+iN)(x_{\A})+ \nn\\&&+
T^{(\mu\alpha)}(x_{\A})]+(\tp)^2(\btp)^2 u^-_k\xi^{\mu k}(x_{\A})\}~.
\lb{comp3}
\eea
Here  $h_a^m,~\psi^{mk}_\alpha$ and $A_a$ are physical fields,
and other components play a role of auxiliary fields.

In Appendix, we consider two bases of $B$-co\-va\-riant differential
operators:
 1) $D=(\dpp,~\dm,~D_{\M})$ in the coordinates $u^\pm_i,~z^{\M}$ and
2) $D^{\A}=(\Dp,~\Dm,~D_{\M}^{\A})$ in the coordinates  \p{B1}.
 Corresponding to these bases decompositions of $G^{\pp}$ and other
objects of differential geometry will automatically  be covariant with
respect to the background supersymmetry.

Define now gravitational  $B$-superfields $G^{\M\pp}$ in the basis
$D_{\M}^{\A}$
\be
\Dep=\Dp + G^{\Hm\pp }\partial_{\Hm}^{\A}
-G^{\hm\pT}D_{\hm}^- ~. \lb{B15}
\ee

 Superfields $G^{\Hm\pp} $  can be written in terms of the analytic
prepotentials of the holonomic basis
\bea
&&G^{m\pp }= h^{m\pp} +2i \tm \sigma^m \bar{h}^{\pT} +
2i  h^{\pT}\sigma^m \btm~, \lb{B22b}\\
&&G^{\5\pp }=h^{\5\pp } -2i (\tm h^{\pT}) +
2i (\btm \bar{h}^{\pT} )~,\lb{B23b}\\
&& G^{\hm\pT}=h^{\hm\pT}~.\lb{B24b}
\eea

By definition,  $B$-superfields have trivial full background $\epsilon$
 variations $\delta_{\B}G^{\M\pp}=0$, and $B$-\-trans\-for\-mations
 of the analytic prepotentials have a noncovariant form
\bea
&&\delta_{\B}h^{m\pp}=2i(\sigma^m)_{\mdn}(\bar{\epsilon}^{k\dot{\nu}}
h^{\mu\pT}-\epsilon^{k\mu}\bar{h}^{\dot{\nu}\pT})u^-_k~, \lb{Btr1}\\
&&\delta_{\B}h^{\5\pp}=2i(\epsilon^{k\mu}
h_\mu^{\pT}+\bar{\epsilon}^{k\dot{\nu}}\bar{h}_{\dot{\nu}}^{\pT})u^-_k~,
 \quad \delta_{\B}h^{\hm\pT}=0~. \lb{Btr2}
\eea

The operator condition of analyticity  $[D^+_{\hm},G^{\pp}]=0$  is
equivalent to the following relations for the gravitational superfields:
\bea
&&D^+_{\hm} G^{\hn\pT}=0~,\quad
D^+_{\hm} G^{\5\pp}=-2i G_{\hm}^{\pT}~,\lb{B16}\\
&&\Dap G^{m\pp }=2i(\sigma^m)_\ada \bar{G}^{\dot{\alpha}\pT}
=-(\sigma^m)_{\ada}\bar{D}^{\dap} G^{\5\pp}~,\lb{B17}\\
&&\bDap G^{m\pp }=-2i(\sigma^m)_\ada G^{\alpha\pT} =(\sigma^m)_\ada
D^{\ap} G^{\5\pp}~.
\lb{B18}
\eea

We shall consider the following decomposition of the transformation
operator $\Lambda$ in covariant bases:
\bea
&&\Lambda =\Lambda^{\M}D_{\M}^{\A}=\hat{\Lambda}^{\M}D_{\M}~,
 \lb{C18}\\
&& \Lambda^m=\hat{\Lambda}^m=\lambda^m + 2i\tm \sigma^m \bar{\lambda}^+
+2i \lambda^+ \sigma^m \btm~,  \lb{C19}\\
&& \Lambda^{\5}=\hat{\Lambda}^{\5}=\lambda^{\5} - 2i(\tm \lambda^+)
+2i (\btm \bar{\lambda}^+)~,\quad\Lambda^{\hm\pm} =\lambda^{\hm\pm}~,
  \lb{C20}\\
&&\hat{\Lambda}^\mu_k=
\lambda^{\mu+}u^-_k -\lambda^{\mu-}u^+_k~.
 \lb{C20c}
\eea

It is evident that these  $B$-su\-per\-field parameters satisfy the
 constraints analogous to eqs.(\ref{B16}-\ref{B18}), in particular, the
 parameters  $\Lambda^{\Hm}$ are linear in $\theta^{\hm-}$.

It should be stressed that the use of background supersymmetry
in the formalism of $N=1$ supergravity leads to appearance of linear
vector superfield parameters which contain chiral parameters in
the  zero and first  orders of the $\bar{\theta}$-decomposition \cite{ZP}.

Transformations of the differential operators of the covariant basis
are
\bea
&&\delta D^{\s\pm\pm}=[\Lambda,D^{\s\pm\pm}]=-(D^{\s\pm\pm}\Lambda^{\M})
D_{\M}^{\A}~,\quad  \delta D_{\M}^{\A}=[\Lambda,D_{\M}^{\A}]\equiv
-\Lambda_{\M}^{\N} D_{\N}^{\A}~,\lb{B25}\\
&&\Lambda_{\5}^{\N}=0~,\quad \Lambda_m^{\Hn}=\partial_m^{\A} \Lambda^{\Hn}~,
\quad\Lambda_{\hm}^{-\hn+}=-D_{\hm}^-\lambda^{\hn+}~,
\quad\Lambda_{\hm}^{+ \hn -}=D_{\hm}^+\lambda^{\hn-}~, \lb{B26}\\
&& \Lambda_{\mu}^{-n}=-D_{\mu}^-\Lambda^n +2i(\sigma^n)_{\mdn}
   \bar{\lambda}^{\dot{\nu}-}~,\quad
   \Lambda_{\mu}^{-\5}=-D_{\mu}^-\Lambda^{\5} -2i\lambda_{\mu}^-~,\nn\\
&& \Lambda_{\dot{\mu}}^{-n}=-\bar{D}_{\dot{\mu}}^-\Lambda^n -
2i(\sigma^n)_{\nu\dot{\mu}} \lambda^{\nu-}~,\quad
   \Lambda_{\dot{\mu}}^{-\5}=-\bar{D}_{\dot{\mu}}^-\Lambda^{\5} -
2i\bar{\lambda}_{\dot{\mu}}^-~.\lb{B27}
\eea

The operator  $\Dep$ is invariant by definition, and this allows
us to derive a transformation law of  $G^{\pp}$  in $\Lambda SG^2_4$
\be
\delta G^{\pp}=-\delta \Dp =[ \Dp, \Lambda ]~. \lb{C17}
\ee

 The full variations of the covariant gravitational superfields have the
following form:
\bea
&&\delta G^{m\pp}=\Dep \Lambda^m +2i(\sigma^m)_{\mdn}(G^{\mu\pT}
\bar{\lambda}^{\dot{\nu}-}-\bar{G}^{\dot{\nu}\pT}\lambda^{\mu-})~,
\lb{C21b} \\
&& \delta G^{\5\pp}=\Dep \Lambda^{\5}-2iG^{\mu\pT}\lambda^-_\mu
-2i\bar{G}^{\dot{\mu}\pT}\bar{\lambda}^-_{\dot{\mu}}~.
\lb{C21c}
\eea

Corresponding local gauge transformations $\hat{\delta}
G^{\M\pp}=\delta G^{\M\pp} -\Lambda G^{\M\pp}$ are
\be
\hat{\delta} G^{\M\pp }=(\Dp  + G^{\pp})\Lambda^{\M}-\lambda G^{\M\pp}
\equiv R^{\M\pp}_{\N}(G)\Lambda^{\N}~,
\lb{trop1}
\ee
where the components of the local transformation operator
$R^{\M\pp}_{\N}(G)$ are defined. It is evident that these transformations
are consistent with the constraints (\ref{B16}-\ref{B18}).

The local variation of the operator $G^{\pp}$ is determined by the
transformations $\hat{\delta} G^{\M\pp}$
\be
\hat{\delta} G^{\pp}=[(\Dp  + G^{\pp}),\Lambda]~, \lb{locoper}
\ee
since the operators $D^{\A}_{\M}$ are not change with respect to local
transformations. This relation is useful
 for calculation of  linear bracket parameters in the commutator of
 transformations  \p{trop1}
\be
\Lambda^{\Hm}(1,2)=\lambda_{\s2}\Lambda^{\Hm}_{\s1}-\lambda_{\s1}
\Lambda^{\Hm}_{\s2}~,\quad \lambda=
\Lambda^n\partial_n-\lambda^{\hn+}D^-_{\hn}~, \lb{brac1}
\ee
which are constructed by analogy with eq.\p{commu} for analytic
 parameters.

The functions  $\lambda^{\hm-}$ in the gauge  \p{B10} are the series
in terms of the superfields $G$
\be
\lambda^{\hm-}(z,u)=\int  \frac{du_{\s1} du_{\s2}}
{(u^+ u_{\s1}^+)}[\lambda^{\hm+}(z,u_{\s1})-
\frac{G^{\pp}(z,u_{\s1})\lambda^{\hm+}(z,u_{\s2})}{(u^+_{\s1}
u_{\s2}^+)}] + O(G^2)~.
\lb{C22}
\ee
Note that the local variation of gravitational superfields (in
distinction with the full variation $\delta G^{\Hm\pp}$ and the operator
$\Lambda$) does not contain $\lambda^{\hm-}$; so the operator
$R^{\M\pp}_{\N}(G)$ in the perturbation theory can be decomposed into
the sum of terms of the first and zero degree in the gravitational
constant (superfields $G$ are linear in $\kappa$).

\setcounter{equation}{0}
\section{\label{H}ALTERNATIVE REPRESENTATIONS  \newline
 OF GRAVITATIONAL SUPERFIELDS}

\hspace{0.5cm}
Let us remind that in the harmonic representation the nonlinear superfield
 constraints of $SG^2_4$ are reduced to the analyticity conditions for the
 prepotentials $H^{\M\pp}$ or to the equivalent linear constraints
(\ref{B16}-\ref{B18}) for the gravitational background superfields.
 Consider the manifestly co\-va\-riant representation
for $G^{\Hm\pp}$\cite{ZT}
\bea
&& G^{m\pp}(\Psi)=(\sigma^m)_{\adb}[(D^+)^2\bar{D}^{\dot{\beta}+}
\Psi^{\alpha-} -(\bar{D}^+)^2D^{\alpha+}\bar{\Psi}^{\dot{\beta}-}]~,
 \lb{D1}\\
&& G^{\5\pp}(\Psi)=(\bar{D}^+)^2D_\alpha^+
   \Psi^{\alpha-} + (D^+)^2\bar{D}_{\dot{\alpha}}^+
\bar{\Psi}^{\dot{\alpha}-}~, \lb{D2}
\eea
which allows one to solve all constraints and to rewrite all geometric
quantities through the unconstrained spinor superfields
$\Psi^-_\alpha(z,u)$ and $\bar{\Psi}^-_{\dot\alpha}(z,u)$ of dimension
$d=5/2$. In a special gauge the spinor prepotentials are linear in
$u^-_k$ and proportional to an ordinary superfield $\Psi^\alpha_k(z)$
analogous to the gauge superfield of the linearized $SG^2_4$  \cite{GS}.

A solution of constraints for the superfield parameters
$\Lambda^{\Hm}$ (\ref{C19},\ref{C20}) can also be written in terms of
 unconstrained parameters $K^{\s(-3)}_\alpha(z,u)$
\bea
&& \Lambda^{m}(K)=(\sigma^m)_{\adb}[(D^+)^2\bar{D}^{\dot{\beta}+}
K^{\alpha\s(-3)} -(\bar{D}^+)^2D^{\alpha+}\bar{K}^{\dot{\beta}\s(-3)}]
~, \lb{param1}\\
&& \Lambda^{\5}(K)=(\bar{D}^+)^2D_\alpha^+K^{\alpha\s(-3)}+
(D^+)^2\bar{D}_{\dot{\alpha}}^+\bar{K}^{\dot{\alpha}\s(-3)}
    ~, \lb{param2}\\
&& \lambda^+_\alpha(K)={i\over 4}(D^+)^2(\bar{D}^+)^2K^{\s(-3)}_\alpha~,
\quad \lambda(K)=\Lambda^{m}(K)\partial_m^{\A} - \lambda^{\hm+}(K)
D^-_{\hm}~.\lb{param3}
\eea

Local transformations of the superfields $\Psi^-_\alpha$ contain the
fermionic parameters $K^{\s(-3)}_\alpha$  as well as the additional
bosonic parameters of transformations
\bea
&&\hat{\delta} \Psi_\alpha^-=[\Dp + G^{\pp}(\Psi)]K^{\s(-3)}_\alpha
-\lambda(K)\Psi_\alpha^- + D^+_\alpha B^{\m}+\nn\\
&&+D^{\beta+} B_{(\alpha\beta)}^{\m}+ \bar{D}^{\dot{\beta}+}
B^{\m}_{\alpha\dot{\beta}}~, \lb{nonge}
\eea
where $B^{\m}$ and $B^{\m}_{\alpha\dot{\beta}}$ are real parameters, and
$B_{(\alpha\beta)}^{\m}$ are symmetrical in spinor indices. It is easy to
show that these transformations produce the local transformations
\p{trop1} for $G^{\Hm\pp}(\Psi)$ dependent on the fermionic
parameters only.

Commutators of  additional transformations vanish, and the bracket
parameters of nontrivial commutators have the following form:
\bea
&&K^{\s(-3)}_\alpha(1,2)=\lambda(K_{\s2})K^{\s(-3)}_{\alpha\s1}
 -(1\leftrightarrow 2)~,\lb{brac}\\
&& B^{\m}_{\hm\hn}(1,2)=\lambda(K_{\s2})B^{\m}_{\hm\hn\s1}-
(1\leftrightarrow 2)~.\lb{brac2}
\eea
A derivation of these relations is given by using eq. \p{locoper}.

By analogy with eqs.(\ref{B22b},\ref{B23b}), we  now consider the
$B$-co\-va\-riant nonlocal representation for the solution of the
$SG^2_4$ constraints (\ref{B16}-\ref{B18}), which is useful in the
analysis of the linearized supergravity
\bea
&&G^{m\pp} =g^{m\pp} +2i(\sigma^m)_{\mdn}[\Theta^{\mu-}
\bar{h}^{\dot{\nu}\pT}
-\bar{\Theta}^{\dot{\nu}-}h^{\mu\pT}]
~,\lb{E4}\\
&&G^{\5\pp} =g^{\5\pp} -2i\Theta^{\mu -}h^{\pT}_\mu-
2i\bar{\Theta}^{\dot{\mu}-}
\bar{h}_{\dot{\mu}}^{\pT}~,\lb{E6}
\eea
where the following operators are introduced:
\be
\Theta^{\mu-}={i\over 2\Box}\partial^{\mdn}\bar{D}_{\dot{\nu}}^-~,\quad
\bar{\Theta}^{\dot{\mu}-}=-{i\over 2\Box}\partial^{\nu\dot{\mu}}D_\nu^-~.
 \lb{E5b}
\ee

It is evident that one can restore locality of the representation using
variables $g^{\pT}_{\hm}=\Box^{-1}h^{\pT}_{\hm}$  of dimension 3/2;
however, this changes dimensions of the corresponding component fields.

It is not difficult to establish relations between the auxiliary
$B$-su\-per\-fields $g$ and the local prepotentials $h$
\bea
&&g^{m\pp}=h^{m\pp}-{1\over\Box}(\sigma^m)_{\mdn}
(\partial^{\mu\dot{\rho}}\bar{\partial}^-_{\dot{\rho}}
\bar{h}^{\dot{\nu}\pT}+\partial^{\rho\dot{\nu}}\partial^-_\rho
 h^{\mu\pT})\lb{nonl1}~,\\
&&g^{\5\pp}=h^{\5\pp}+{1\over\Box}
(\partial^{\mu\dot{\rho}}\bar{\partial}^-_{\dot{\rho}}h_\mu^{\pT}
-\partial^{\rho\dot{\mu}}\partial^-_\rho \bar{h}_{\dot{\mu}}^{\pT})
\lb{nonl2}~.
\eea

Due to the nonlocality of these relations, one should carefully study
connections between field components of these superfield representations.
In various treatments of the gauge symmetry components of the superfields
$g^{\M\pp}$ could differ from the standard set of components of the
 prepotentials $h^{\M\pp}$ (\ref{comp1}-\ref{comp3}).

Consider the additive gauge transformations of the superfields
$g^{\Hm\pp}$ induced by linearized transformations of $h^{\Hm\pp}$
\bea
&&\delta_{\0}g^{m\pp}=\Dp l^{m}= \Dp [\lambda^{m}-
{1\over\Box}(\sigma^m)_{\mdn}(\partial^{\mu\dot{\rho}}
\bar{\partial}^-_{\dot{\rho}}\bar{\lambda}^{\dot{\nu}+}
+\partial^{\rho\dot{\nu}}\partial^-_\rho \lambda^{\mu+})\lb{nonltr1}]~,\\
&&\delta_{\0}g^{\5\pp}=\Dp l^{\5}= \Dp [\lambda^{\5}-
{1\over\Box}
(\partial^{\mu\dot{\rho}}\bar{\partial}^-_{\dot{\rho}}\lambda_\mu^{+}
-\partial^{\rho\dot{\mu}}\partial^-_\rho \bar{\lambda}_{\dot{\mu}}^{+})]
\lb{nonltr2}~.
\eea

The transformations are nonlocal in this treatment,
and components of $g^{\Hm\pp}$ are nonlocal combinations of the
standard components. If one would treat $g^{\Hm\pp}$ and the parameters
$l^{\Hm}$ as independent analytic superfields,  a stuff of auxiliary
components could change essentially; so we shall not develop here such a
treatment.

\setcounter{equation}{0}
\section{\label{C} SOLUTION OF HARMONIC CONSTRAINTS IN  $SG^2_4$ }

\hspace{0.5cm}
Solution of the basic condition of the Grassmann analyticity allows one
to construct all geometric quantities of $SG^2_4$: supervielbein,
 connection and tensors of torsion and curvature.

In the harmonic formalism of  $SG^2_4$, one introduces the 2-nd invariant
harmonic operator in addition to the basic operator \p{B15}
\be
\Dem=\Dm + G^{\m}~,\quad G^{\m}=h^{\M\m}\partial_{\M}^{\A}=G^{\M\m}
D_{\M}^{\A}~,
\lb{harmin}
\ee
where the corresponding coefficients are considered in different bases.
The $B$-su\-per\-fields  $G^{\Hm\m},G^{\hm-}$ and $G^{\hm\s(-3)}$ play an
important role in the geometry of supergravity; they can be written
through the basic superfields $G^{\Hm\pp}$ in  perturbation theory.

Gauge transformations of the $B$-su\-per\-fields $G^{\M\m}$ are determined
 by the invariance condition of the operator $\Dem$
\bea
&&\delta G^{m\m}=\Dem \Lambda^m +2i(\sigma^m)_{\mdn}(G^{\mu-}
\bar{\lambda}^{\dot{\nu}-}-\bar{G}^{\dot{\nu}-}\lambda^{\mu-})~,
\lb{Cm1} \\
&& \delta G^{\5\m}=\Dem \Lambda^{\5}-2iG^{\mu-}\lambda^-_\mu
-2i\bar{G}^{\dot{\mu}-}\bar{\lambda}^-_{\dot{\mu}}~,
\lb{Cm2}\\
&&\delta G^{\hm-}=\Dem \lambda^{\hm+}- \lambda^{\hm-}~,\quad
\delta G^{\hm\s(-3)}=\Dem \lambda^{\hm-}~.\lb{Cm3}
\eea

Commutation relation between the invariant harmonic derivatives
is the fundamental constraint of the harmonic formalism of
$SG^2_4$  \cite{GIK1,Z1}
\be
[\Dep ,\Dem ]=[(\Dp+G^{\pp}),(\Dm+G^{\m})]=D^{\0}~.
 \lb{C2}
\ee

This harmonic equation has the manifest iterative solution \cite{Z1}
\be
G^{\m}=\sum\limits^{\infty}_{n=1} (-1)^n \int du_1 \ldots
du_n \frac{G^{\pp}(z,u_{\s1})\ldots G^{\pp}(z,u_n )}{(u^+ u^+_{\s1})
\ldots (u^+_n u^+ )}~.  \label{C4}
\ee
A solution to the functions $h^{\M\m}$ has an analogous structure in all
degrees of perturbation theory.

$B$-covariant equations for the coefficients in expansion of this operator
have the following form:
\bea
&&\Dp G^{m\m }-\Dm G^{m\pp } + G^{\pp}G^{m\m }-G^{\m}G^{m\pp }+\nn\\
&&+ 2i(\sigma^m)_{\mdn }(G^{\mu\pT} \bar{G}^{\dot{\nu}\s(-3)}+
G^{\mu\s(-3)} \bar{G}^{\dot{\nu}\pT})=0~, \lb{C6}\\
&&\Dp G^{\5\m }-\Dm G^{\5\pp } + G^{\pp}G^{\5\m }-G^{\m}G^{\5\pp }-\nn\\
&&- 2iG^{\mu\pT} G^{\s(-3)}_\mu +2i\bar{G}_{\dot{\mu}}^{\pT}
\bar{G}^{\dot{\mu}\s(-3)}=0~, \lb{C7}\\
&&\Dp G^{\hm -}-\Dm G^{\hm\pT}
+ G^{\pp}G^{\hm-}-
G^{\m}G^{\hm\pT}=0~, \lb{C8}\\
&& G^{\hm -}=(\Dp +G^{\pp})G^{\hm\s(-3)}\lb{C8b}~.
\eea

These equations can readily be solved  in  perturbation theory; for
instance, in the first two orders we have
\bea
&&G^{\Hm\m}_{\s(1)}(z,u)=\int \frac{du_{\s1} }{(u^+ u_{\s1}^+)^2}
                        G^{\Hm\pp}(z,u_{\s1})~, \lb{Cv1}\\
&& G^{\hm\s(-2\pm 1)}_{\s(1)}(z,u)=\int \frac{du_{\s1}(u^\pm u_{\s1}^-)}
{(u^+ u_{\s1}^+)^2}G^{\hm\pT}(z,u_{\s1})~, \lb{Cs1}\\
&&G^{m\m}_{\s(2)}(z,u)=\int \frac{du_{\s1} du_{\s2}}{(u^+ u_{\s1}^+)
(u_{\s1}^+ u_{\s2}^+)(u_{\s2}^+ u^+)} \{G^{\pp}(z,u_{\s1})
G^{m\pp}(z,u_{\s2})+\nn\\
&&+2i(u_{\s1}^- u_{\s2}^-)(\sigma^m)_{\mdn}G^{\mu\pT}(z,u_{\s1})
\bar{G}^{\dot{\nu}\pT}(z,u_{\s2})\}~,
\lb{Cv2}\\
&&G^{\5\m }_{\s(2)}(z,u)=\int \frac{du_{\s1} du_{\s2}}{(u^+ u_{\s1}^+)
(u_{\s1}^+ u_{\s2}^+)(u_{\s2}^+ u^+)} \{G^{\pp}(z,u_{\s1})
G^{\5\pp}(z,u_{\s2})-\nn\\
&&-i(u_{\s1}^- u_{\s2}^-)[G^{\mu\pT}(z,u_{\s1}) G^{\pT}_\mu(z,u_{\s2})-
\bar{G}_{\dot{\mu}}^{\pT}(z,u_{\s1})\bar{G}^{\dot{\mu}\pT}(z,u_{\s2})]\}~,
\lb{C9}
\eea
where the harmonic distributions $(u_{\s1}^{k\pm} u_{k\s2}^\pm)$
and  $(u^{k+}_{\s1}u^+_{k\s2})^{-n}$ \cite{GI2} are considered.

It is reasonable to discuss the 1-st order solution only in the nonlocal
representation (\ref{E4}-\ref{E6})
\bea
G^{m\m}_{\s(1)}=g^{m\m}_{\s(1)}+2i(\sigma^m)_{\mdn}[\Theta^{\mu-}
\bar{h}^{\dot{\nu}-}_{\s(1)}-\bar{\Theta}^{\dot{\nu}-}h^{\mu-}_{\s(1)}
-\Theta^{\mu+}\bar{h}^{\dot{\nu}\s(-3)}_{\s(1)}+\bar{\Theta}^{\dot{\nu}+}
h^{\mu\s(-3)}_{\s(1)}]~,&&\lb{nlmin1}\\
G^{\5\m} =g^{\5\m}_{\s(1)} +2i[\Theta_\mu^- h^{\mu-}_{\s(1)}
+\bar{\Theta}_{\dot{\mu}}^-\bar{h}_{\s(1)}^{\dot{\mu}-}-
\Theta_{\mu}^+ h^{\mu\s(-3)}_{\s(1)}
 -\bar{\Theta}_{\dot{\mu}}^+\bar{h}_{\s(1)}^{\dot{\mu}\s(-3)}]
~,&&\lb{nlmin2}
\eea
where  $\Theta^{\mu+}={i\over 2\Box}\partial^{\mu\dot{\nu}}
D^+_{\dot{\nu}}$, and also there are defined quantities
$g^{\Hm\m}_{\s(1)}$ and $h^{\hm\s(-2\pm 1)}_{\s(1)}$ which can be
written via the corresponding prepotentials by analogy with
(\ref{Cv1},\ref{Cs1}).

\setcounter{equation}{0}
\section{SUPERFIELD DECOMPOSITIONS \newline
OF SUPERVIELBEIN AND CONNECTION}

\hspace{0.5cm}
Differential geometry of $SG^2_4$ has been considered in the holonomic
basis \cite{GNS}. We shall study the background supersymmetry of basic
geometric objects in the $B$-co\-va\-riant basis. By analogy with the
formalism of $D=4, N=1$ conformal supergravity \cite{Z5} one can
introduce in $HSS$ a so-called {\it almost covariant} basis of
differential operators $E_{\B}$, which helps to define $B$-su\-per\-field
blocks necessary for construction of supervielbeins and connections of
the theory. The initial step of this construction is connected
with the following spinor operator:
\be
E^-_{\ha}\equiv[D^+_{\ha} , \Dem]=-D^-_{\ha}+ [D^+_{\ha},G^{\m}]~.
\lb{C1}
\ee

Define also vector and scalar operators
\bea
&&E_{a}=-{i\over 4}(\bar{\sigma}_a)^{\dot{\beta}\alpha}\{D^+_\alpha,
E^-_{\dot{\beta}}\}=\partial_a +{i\over 4}(\bar{\sigma}_a)^{\dot{\beta}
\alpha}\{D^+_\alpha,[G^{\m},\bar{D}^+_{\dot{\beta}}]\}~,\lb{C1b}\\
&&E_{\5}={i\over 2}\{D^{\alpha+},E^-_\alpha\}=\partial_{\5}
-{i\over 2}\{D^{\alpha+},[G^{\m},D^+_\alpha]\}~,\lb{C1c}\\
&&\bar{E}_{\5}={i\over 2}\{\bar{D}^{\dot{\alpha}+},E^-_{\dot{\alpha}}\}
~.\lb{C1d}
\eea

By definition,  components of the basis $E_{\B}$ satisfy the relations
\be
[\Dep,E^-_{\ha}]=-D^+_{\ha}~,\quad [\Dem,E^-_{\ha}]= 0~,\quad
[\Delta^{\s\pm\pm}, E_{a}]=[\Delta^{\s\pm\pm}, E_{\5}] =
[\Delta^{\s\pm\pm},\bar{E}_{\5}]=0~. \lb{har}
\ee

Decomposition of the almost covariant operator
$E_{\B}=G_{\B}^{\M} D_{\M}^{\A}$ determines matrix elements of
supervielbein. The corresponding density $E=\mbox{Ber}G_{\B}^{\M}$ has
 the correct transformation law
\be
\delta E=(\partial_m \Lambda^m +D^-_{\hm}\lambda^{\hm+}-
D^+_{\hm}\lambda^{\hm-})E~. \lb{dens}
\ee
Note that the density  $E$ of this theory is defined uniquely and does not
depend on a choice of basis; however, we prefer to use its expression in
terms of the $B$-su\-per\-fields.

By definition, this quantity is $B$-covariant and does not depend on the
scalar superfield $G^{\5\pp}$. The linear approximation for $E$ is
\be
E_{\s(1)}=D^+_{\hm}G^{\hm -}_{\s(1)} - {i\over 4}D^{\alpha+}
\bar{D}^{\dot{\beta}+}G^{\m}_{\alpha\dot{\beta}\s(1)}~. \lb{C30}
\ee

Higher-order terms in  $E$ are calculated straightforwardly from
decompositions of the supervielbein or by using the equation
\be
[\Dp+G^{\pp}-\partial_m^{\A}G^{m\pp}-D^-_{\hm}G^{\hm\pT}]E=0~.
\lb{deneq}
\ee

A gauge-covariant  $SL(2,C)$-basis in the harmonic superspace of
$SG^2_4$ has been constructed in ref.\cite{GNS}. One defines the
$SL(2,C)$-co\-va\-riant spinor operator instead of the flat operator
 $D^+_\alpha$
\be
\Delta^+_\alpha\equiv u^+_i\Delta^i_\alpha=D^+_\alpha +
F_\alpha^{\dot{\mu}}\bar{D}^+_{\dot{\mu}}~,
 \lb{C10}
\ee
where
\be
 F_\alpha^{\dot{\mu}}=\Delta^+_\alpha\bar{\psi}^{\dot{\mu}-}=
D^+_\alpha\bar{\psi}^{\dot{\nu}-}(1-\bar{D}^+
\bar{\psi}^-)^{-1\dot{\mu}}_{\dot{\nu}}~, \lb{C10b}
\ee
is the matrix which is expressed in terms of the transition function
 $\bar{\psi}^{\dot{\mu}-}$ for the so-called hybrid basis.

We consider the terms in this matrix linear in $\kappa$
\be
F_{\alpha\dot{\mu}\s(1)}=-D^+_\alpha \bar{G}_{\dot{\mu}\s(1)}^- +
\frac{i}{8} (\sigma_n)_{\alpha\dot{\mu}}(D^+)^2 G^{n\m }_{\s(1)}=
D^+_\alpha\bar{\psi}_{\dot{\mu}\s(1)}^-~. \lb{C15}
\ee

The $SL(2,C)$-basis also contains  a function $F$ which depends on the
superfield $G^{\5\pp}$ \cite{GNS}. The linear approximation for this
function is
\bea
&&F_{\s(1)}=-{1\over 2} D_\mu^+ G^{\mu-}_{\s(1)}-{i\over 8}(D^+)^2
G^{\5\m}_{\s(1)}~,\lb{C25}\\
&&\hat{\delta}_{\s(0)}F_{\s(1)}={1\over 2} D^+_\mu\lambda^{\mu-}~.
 \lb{C26}
\eea
The functions $F,\bar{F},F_\alpha^{\dot{\mu}}$ and
$\bar{F}_{\dot{\alpha}}^\mu$ determine components of superfield
connection.

\setcounter{equation}{0}
\section{\lb{D} $B$-SUPERFIELD ACTION OF $SG^2_4$ }

\hspace{0.5cm}
In the harmonic $SG^2_4$ formalism we can consider iterative construction
of the $B$-su\-per\-field action $ S^{cl}=\sum S_{\s(n)}^{cl}$ which
begins with the quadratic action of linearized theory $S_{\s(2)}^{cl}$,
and terms of higher order in superfields $G$ should be restored with the
help of the gauge symmetry
\be
\hat{\delta}_{\s(1)} S_{\s(n)}^{cl} + \hat{\delta}_{\s(0)}
S_{\s(n+1)}^{cl}=0~. \lb{D2b}
\ee

The linearized harmonic action has been constructed in our work \cite{ZT}
\be
S_{\s(2)}^{cl}={1\over 2\kappa^2}\int d^{\s12}z du
[G^{\5\pp}G^{\5\m}_{\s(1)}+{1\over 2}G^{m\pp}G^{\m}_{m\s(1)}]~,
\lb{D3}
\ee
where eq. \p{Cv1} is used. This expression is a quadratic form,  each
term being similar to the action of the abelian gauge superfield
$V^{\pp}$ in the full harmonic superspace. It should be noted that this
quadratic form possesses  additional $SO(3,2)$ symmetry, and the
components of the 5-vector $G^{\Hm\pp}$ have a nonstandard normalization
with respect to this symmetry.

One can easily verify, using $WZ$-gauge (\ref{comp1}-\ref{comp3}),
that this action is equivalent to the component action of linearized
$SG^2_4$. This can be checked quite simply for the terms quadratic in
auxiliary components . The corresponding terms in $G^{\Hm\m}_{\s(1)WZ}$
can be calculated in the coordinates $z^{\M}_{\A}$ using  relations of
the following type:
\bea
\Dm[(\tp)^2\btpa u^-_k]=\Dp[(\tp\tm)\btma u^-_k+{1\over 2}(\tm)^2\btpa
u^-_k -{1\over 2}(\tm)^2\btma u^+_k]~,&&\lb{ancor1}\\
 \Dm[(\tp)^2\tma \btpa]=-\Dp[(\tm)^2\tpa\btma]~.\lb{ancor2}&&
\eea

It is useful to consider the equivalent analytic representation of
  $S_{\s(2)}^{cl}$
\be
{1\over 2\kappa^2}\int \dza du [G^{\5\pp}T^{\5\pp}_{\s(1)}
+{1\over 2}G^{m\pp}T^{\pp}_{m\s(1)} + G^{\hm\pT}T^+_{\hm\s(1)}]~, \lb{D5}
\ee
where we introduce the linearized components of torsion which can easily
be written via spinor derivative of harmonic superfields
$G^{\Hm\m}_{\s(1)}$  comparing  various integral representations of
$S_{\s(2)}$ and using eqs.\p{G13}.
\bea
&&T^+_{\mu\s(1)}=-{i\over 4}(\bar{D}^+)^2D^+_\mu G^{\5\m}_{\s(1)}-
{i\over 8}(D^+)^2\bar{D}^{\dot{\nu}+} G^{\m}_{\mu\dot{\nu}\s(1)}~,
\lb{D6}\\
&& D^+_\nu T^+_{\mu\s(1)}=-2i\varepsilon_{\nu\mu}T^{\5\pp}_{\s(1)}~,
\quad \bar{D}^+_{\dot{\nu}} T^+_{\mu\s(1)}=
-iT^{\pp}_{\mu\dot{\nu}\s(1)}~.\lb{D8}
\eea

Note that the linearized components of torsion satisfy the following
conditions:
\be
\Dp T^+_{\hm\s(1)}=0~,\quad D^+_{\ha}D^+_{\hb}T^+_{\hm\s(1)}=0~.
\lb{subcon}
\ee

The action $S_{\s(2)}$ is invariant under the linearized transformations
$\hat{\delta}_{\s(0)}G^{\pp}_{\Hm}=\Dp \Lambda_{\Hm}$ .

Using the representation
(\ref{D1},\ref{D2}) for the superfields $G^{\Hm\pp}(\Psi)$ in the action
\p{D3} one can obtain the linearized equation of motion for
 $SG^2_4$ varying in the spinor prepotential $\Psi^{\mu-}$
\be
T^+_{\mu\s(1)}(\Psi)=0~. \lb{D20}
\ee

In the nonlocal superfield representation  (\ref{E4},\ref{E6}), one
can construct the invariant of linearized transformations for the
analytic $B$-su\-per\-field $g^{\5\pp}$
\be
\int \frac{d^{\s12}z du_{\s1}du_{\s2}}
{(u^+_{\s1}u^+_{\s2})^2}
g^{\5\pp}(u_{\s1})g^{\5\pp}(u_{\s2})
\lb{actnl}
\ee
and the analogous independent invariant for $g^{m\pp}$. The only  local
 invariant of local transformations and appropriate dimension exists in
the initial superfield variables \p{D3}.

The nonlocal invariant  can also be constructed in terms
of the gauge spinor superfield
\be
\int \frac{d^{\s12}z du_{\s1}du_{\s2}}{\Box(u^+_{\s1}u^+_{\s2})^3}
h^{\mu\pT}[\pmn -{i\over 2}(u^+_{\s1}u^+_{\s2})D^-_{\mu\s1}
\bar{D}^-_{\dot{\nu}\s2}]\bar{h}^{\dot{\nu}\pT}~. \lb{spin}
\ee
In principle, nonlocal invariants can be used in  the quantum
effective action of this theory.

Note that it is possible to use an alternative approach in which the
background superspace is connected with a flat supersymmetry without
the central charge $B^2_4\equiv B^\prime$
\be
\Dep=\Dp_{\s4} + G^{m\pp }\partial_m^{\A}+H^{\5\pp}\partial_{\5}^{\A}
-G^{\hat{\mu}\pT}(D_{\s4}^-)_{\hat{\mu}}~, \lb{C32}
\ee
where the
$B^\prime$-covariant harmonic and spinor derivatives $\Dp_{\s4}$ and
$D_{\s4}^-$ (independent of derivative $\partial_{\5}^{\A}$) are
introduced. Transformations of the superfields $G^{m\pp}$ and
 $G^{\hat{\mu}\pT}$ are identical in different background bases, but
the analytic compensator transforms covariantly with respect to the
$B^\prime$ supersymmetry only, and its transformations in $B^2_4(Z)$
contain inhomogeneous $\theta$-de\-pen\-dent terms. Note that connection
of different harmonic superfield representations and background bases has
been discussed, for the gauge theory, in refs. \cite{Z2,BBIKO,IKZ}.

A nonlinear action of $SG^2_4$ в работе has been constructed in ref.
\cite{GNS} as the gauge-invariant action of analytic compensator \p{C32}
\be
S(H^{\5})=\int d^{\s12}z du E^{-1}H^{\5\pp}H^{\5\m}~,\lb{D21}
\ee
 which possesses  independent $B^\prime$-invariance by definition.
The use of the expression of $H^{\5\s\pm\pm}$ via the $B$-su\-per\-fields
$G^{\Hm\pp}$  in this formula leads to the appearance of terms manifestly
depending on spinor coordinates, in particular, the corresponding
quadratic action contains similar terms. Apparently, it is possible to
reconstruct this action as an expansion in terms of superfields $G$ and
see a transmutation of the background supersymmetry, but we prefer simple
iterative constructions with the manifest $B$-co\-va\-riance.

\setcounter{equation}{0}
\section{\lb{F} ON DEVELOPMENT OF THE HARMONIC \newline
FORMALISM OF $SG^2_4$}

\hspace{0.5cm} Our investigation can be used for the superfield
quantization of $N=2$ supergravity in the flat background superspace by
analogy with the superfield formalism of quantization in $N=1$
supergravity \cite{GGRS,ZP}. The use of the representation of
gravitational superfields via the spinor harmonic superfield
$\Psi^-_\alpha(z,u)$ seems to be most adequate for the solution of this
problem. A more interesting and difficult problem is the study of
nonperturbative structure of theory taking into account results
that can be obtained by perturbative superfield methods and by the dual
transformations, insufficiently analyzed in supergravity.
 We shall discuss one possibility of a dual description of $N=2$
supergravity in terms of alternative superfield variables in the harmonic
 approach.

Superfield constraints of the nonlinear $SG^2_4$ theory are reduced to the
kinematic analyticity condition in the standard formalism , and nonlinear
equation of motion for the superfields $G^{\M\pp}$ are  additional
dynamical restrictions on the components of torsion.

By analogy with a dual formulation of the $N=2$ supersymmetric gauge
theory \cite{Z3} we can consider the dual harmonic formalism of the
linearized $SG^2_4$, in which the equation for the spinor ($d=-1/2$)
component of torsion is solved, and then the dynamical analyticity
condition arises. The basic operator of the dual formalism is
$\Dem=\Dm + G^{\m}$, and the linearized equation of motion of the standard
approach with the analytic $G^{\pp}$ transforms into a solvable linear
 constraint $T^+_{\mu\s(1)}=0$ for the dual superfields $G^{\M\m}$.
A solution of this constraint can be written in terms of the new
nonanalytic spinor prepotentials $A^{\hm\s(-3)}$ of dimension 3/2
\bea
&& G^{\5\m}(A)=D^+_\mu A^{\mu\s(-3)} + \bar{D}^+_{\dot{\mu}}
\bar{A}^{\dot{\mu}\s(-3)}~, \lb{F1}\\
&& G^{\m}_{\mdn}(A)= \bar{D}_{\dot{\nu}}^+ A_\mu^{\s(-3)}-D^+_\mu
\bar{A}_{\dot{\nu}}^{\s(-3)}~, \lb{F2}\\
&&T^+_{\mu\s(1)}(A)\sim (\bar{D}^+)^2D_\mu^+ G^{\5\m}(A)+{1\over 2}
(D^+)^2\bar{D}^{\dot\nu+}G_{\mu\dot{\nu}}^{\m}(A)\equiv 0~.
\eea

The operator $G^{\pp}$ in this formalism can be calculated with the help
of eq.\p{C2} and does not satisfy the analyticity condition off-shell.
The dynamical zero-curvature condition becomes a dual equation of motion
\be
[D^+_{\ha} , \Dep ]=0 \lb{F3}~,
\ee
which is equivalent to the linearized equations of motion for the physical
fields of $SG^2_4$ and to the disappearance of all auxiliary components.

Equations of the dual formalism of $SG^2_4$ for the prepotentials
 $A_{\hm}^{\s(-3)}$ are equivalent to  standard equations for the linear
superfields $G$; however, off-shell structures are essentially different
in alternative formulations. In particular, the dual formalism has an
infinite number of auxiliary components in the gravitational multiplets
which vanish on the mass shell only.

I am grateful to E.A. Ivanov for  interesting discussions.
This work is partially supported  by  grants  RFBR-96-02-17634,
RFBR-DFG-96-02-00180,  INTAS-93-127-ext and INTAS-96-0308, and
by  grant of Uzbek Foundation of Basic Research N 11/97.

\setcounter{equation}{0}
\section{\lb{G} APPENDIX }

\hspace{0.5cm} In this appendix we shall consider some useful definitions
and relations connected with the flat harmonic superspace
\cite{GIK1}. The harmonic derivatives $\dpp,~\dm$ and
$\partial^{\0}$ satisfy the relations of the Lie algebra $SU(2)$
and are defined by their action on the harmonics
\bea
&& [\dpp,\dm]=\partial^{\0}~,\quad [\partial^{\0},\partial^{\s\pm\pm}]=
\pm 2\partial^{\s\pm\pm}~, \lb{G1}\\
 && \dpp\;u^+_i=0,\quad \dpp\;u^-_i=u^+_i,\quad \partial^{\0}u^\pm_i=
\pm u^\pm_i~,\lb{G2} \\
  &&\dm\;u^-_i=0,\quad \dm\;\;u^+_i=u^-_i~. \lb{G3}
\eea

The holonomic basis in the set of differential operators
 $\partial^{\A}$ contains partial derivatives on the coordinates
  $z^{\M}_{\A}$ \p{B1}
\be
\partial^{\A}=(\dpp,~\dm,~\partial^{\A}_m,~\partial^{\A}_{\5},~
\partial^-_\mu,~\bar{\partial}^-_{\dot{\mu}},~\partial^+_\mu,~
\bar{\partial}^+_{\dot{\mu}})~.\lb{G4}
\ee
The holonomic basis in the flat real coordinates
  $z^{\M}$  consists of the operators $\partial/\partial z^{\M} $.

 The $B_4^2(Z)$-covariant harmonic derivatives have the following form:
\bea
\Dp=\dpp -2i\tpm\btpn\pmn^{\A} + i[(\tp)^2-(\btp)^2]
\partial_{\5}^{\A} + \tpm\partial^+_\mu +
\btpm\bar{\partial}^+_{\dot{\mu}}~, \lb{G5}\\
\Dm=\dm -2i\tmm\btmn\pmn^{\A} + i[(\tm)^2-(\btm)^2]
\partial_{\5}^{\A} + \tmm\partial^-_\mu +
\btmm\bar{\partial}^-_{\dot{\mu}}~,&&\lb{G6}
\eea
where $(\theta^\pm)^2=\theta^{\alpha\pm}\theta^\pm_\alpha$ and
$(\bar{\theta}^\pm)^2=\bar{\theta}^\pm_{\dot{\beta}}
\bar{\theta}^{\dot{\beta}\pm}$.

Write also known expressions for $B$-covariant spinor derivatives
\bea
&&D^-_\mu=-\partial/\partial \tpm + 2i\btmn\pmn^{\s A}
-2i\tm_\mu\partial^{\A}_{\5}~, \lb{G7}\\
&&\bar{D}^-_{\dot{\nu}}=
  - \partial/\partial\btpn -2i\tmm \pmn^{\s A}
-2i\btm_{\dot{\nu}}\partial^{\A}_{\5}~,\lb{G8}\\
&& D^+_\mu=\partial/\partial \tpm \equiv \partial^+_\mu~,
\quad \bar{D}^+_{\dot{\mu}}=\partial/\partial\btpm \equiv
   \bar{\partial}^+_{\dot{\mu}}~.\lb{G9}
\eea

These operators form the $B$-covariant basis
\be
D^{\A}=(\Dp,~\Dm,~\partial^{\A}_m,~\partial^{\A}_{\5},~
-D^-_{\hm},~D^+_{\hm})~.\lb{G10}
\ee

The alternative $B$-covariant basis is independent of harmonics
\be
D_{\M}=(\partial_m,~\partial_{\5},~D^k_{\hm})~,\quad
[D_{\M} , D^{\s\pm\pm}]=0 \lb{G10b}
\ee
and contains covariant spinor derivatives in the coordinates
 $z^{\M}$
\be
D^k_\mu=\partial^k_\mu + i\bar{\theta}^{k\dot{\nu}}\pmn -
i\theta^k_\mu\partial_{\5}~,\quad \bar{D}^k_{\dot{\mu}}=
\bar{\partial}^k_{\dot{\mu}}-i\theta^{\nu k}\partial_{\nu\dot{\mu}}
-i\bar{\theta}^k_{\dot{\mu}}\partial_{\5}~. \lb{G10c}
\ee

We shall use the following condensed notation for scalar elements in the
algebra of spinor derivatives:
\bea
&&(D^\pm)^2=D^{\alpha\pm}D^\pm_\alpha~,\quad (\bar{D}^\pm)^2=
\bar{D}^\pm_{\dot{\beta}}\bar{D}^{\dot{\beta}\pm}~,\lb{G11}\\
&&(D^\pm)^4= \frac{1}{16}(D^\pm)^2(\bar{D}^\pm)^2~,\quad (D)^4=
\frac{1}{16}(D^+)^2(D^-)^2~. \lb{G12}
\eea
These elements are included in the definition of integration measure
of the full harmonic superspace and the analytic measure
\be
d^{\s12}z_{\A}=d^{\s4}x_{\A}(D^-)^4(D^+)^4~,\quad d^{\s12}z=d^{\s4}x
(D)^4(\bar{D})^4~,\quad\dza =d^{\s4}x (D^-)^4~.
\lb{G13}
\ee

\end{document}